\documentclass[superscriptaddress,longbibliography]{revtex4-2}
\usepackage{graphicx}

\graphicspath{images/}
\usepackage[utf8]{inputenc}
\usepackage{import}
\usepackage{subfiles}
\usepackage[justification=raggedright,singlelinecheck=false]{caption}
\usepackage{pgfplots}
\usepackage{todonotes}
\usetikzlibrary{shapes.arrows}
\usetikzlibrary{calc}
\usepackage[mode=buildnew]{standalone}
\usepackage{tabularx}
\usepackage{makecell}
\usepackage{multirow}
\usepackage{physics}
\usepackage{comment}
\usepackage{float}
\floatstyle{plaintop}
\restylefloat{table}

\let\oldcite\cite
\renewcommand*\cite[1]{~\oldcite{#1}}
\def\suppref{} 

\begin{document}

\title{The Chlorine Vacancy in 4H-SiC: An NV-like Defect With Telecom Emission}

\author{Oscar Bulancea-Lindvall} 
\affiliation{Department of Physics, Chemistry and Biology, Link\"oping
  University, SE-581 83 Link\"oping, Sweden}

\author{Joel Davidsson} 
\affiliation{Department of Physics, Chemistry and Biology, Link\"oping
  University, SE-581 83 Link\"oping, Sweden}

\author{Rickard Armiento} 
\affiliation{Department of Physics, Chemistry and Biology, Link\"oping
  University, SE-581 83 Link\"oping, Sweden}

\author{Igor A. Abrikosov} 
\affiliation{Department of Physics, Chemistry and Biology, Link\"oping
  University, SE-581 83 Link\"oping, Sweden}

\date{\today}

\begin{abstract}
    The diamond nitrogen vacancy (NV) center remains an ever increasing topic of interest. At present, it is considered an ideal example of a solid-state qubit applicable in quantum communication, computing, and sensing alike. With its success, the search for defects that share or improve upon its advantageous features is an ongoing endeavor. By performing large-scale high-throughput screening of 52\,600 defects in 4H silicon carbide (SiC), we identify a collection of NV-like color-centers of particular interest. From this list, the single most promising candidate consists of a silicon vacancy and chlorine substituted on the carbon site, and is given the name of the chlorine vacancy (ClV) center. Through high-accuracy first-principle calculations, we confirm that the ClV center is similar to the NV center in diamond in its local structure and shares many qualitative and quantitative features in the electronic structure and spin properties. In contrast to the NV center, however, the ClV center in SiC exhibits emission in the telecom range near the C-band.
\end{abstract}
\maketitle

\section{Introduction}

Point defects in semiconductors have been studied for more than a decade for applications within quantum technology. They exhibit extraordinarily stable spin states that are optically addressable even at room temperature\cite{Herbschleb2019,Anderson2022,Koehl2011,Wang2020}, feature spin-photon entanglement for quantum communication purposes\cite{Wei2022,Togan2010,Hensen2015,Bernien2013}, and enable delicate sensing of external fields on the nanoscale\cite{Rondin2014,Maertz2010,Pham2011,Mzyk2022,Wickenbrock2016,ZFSSensorPaper,Kucsko2013,Trusheim2016}. The centerpiece of this development has been the nitrogen-vacancy (NV) center in diamond, covering applications in all mentioned categories. Its photoluminescence in the visible range\cite{Gali2019} and spin-dependent electron transition processes have been utilized to achieve spin-photon entanglement for long-distance quantum communication\cite{Hensen2015}. Spin-dependent biases within the radiative and nonradiative deexcitation processes of the NV center have yielded effective and fully optical spin initialization and readout methods\cite{Gali2019}. NV spin-qubit registers utilizing nearby hyperfine-coupled lattice spins have been demonstrated and are of interest in the design of solid-state quantum processors and memories\cite{NVCenterQubit,Bradley2019,Gonzalez2022}, in particular, due to the long spin coherence time of the neighboring nuclear spins, such as the ${}^{14}$N spin. Various sensing methods have been achieved with the NV center, based on altering its spin properties under a wide collection of external fields. In particular, scalar and vector-field magnetometry\cite{Rondin2014,Maertz2010,Pham2011}, temperature\cite{Kucsko2013}, and strain\cite{Trusheim2016} sensing via the alteration of spin levels under the corresponding fields are enabled by optically detected and electron spin resonance (ODMR and ESR). Sensing techniques based on relaxometry near spin-level anti-crossings have also been demonstrated\cite{Mzyk2022}, with the benefit of being microwave-free. On the other hand, the visible range emission is a disadvantage within modern fiber-optical networks. The propagation loss of visible light in optical fibers is considerable, limiting both the viable signal distance and spin-photon entanglement rate\cite{Dreau2018,Hensen2015}. Down-conversion of the photons has been applied to alleviate this issue\cite{Dreau2018}, yielding C-band telecom wavelengths (ca 1550 nm) where propagation loss is minimal. However, down-conversion inevitably decreases the signal-to-noise ratio\cite{Dreau2018}, at times by several orders of magnitude, which leaves an inherent telecom emission a particularly desired feature.

The success and ongoing developments of the diamond NV center continue to inspire exploration of other defects for further improved qualities in hosts offering additional advantages compared to diamond. Silicon carbide (SiC) is one such host with a long history within high-power electronics and electronic device integration, featuring multiple polytypes for numerous defect variations. It offers increased thermal conductivity and mature nanofabrication processes, while its color centers show long coherence times and strong optical emissions\cite{Majety2022}. Following the properties of the diamond NV center, the SiC divacancy and corresponding NV center have garnered attention for their similarity to the diamond NV center in electronic structure, but with emission in the infrared reducing the need for down-conversion\cite{Christle2017,Zargaleh2016}. Combining dynamical decoupling schemes with recent techniques for increased readout fidelity and integration-time, e.g. spin-to-charge conversion, the divacancy has displayed coherence times of more than five seconds in isotope-purified samples\cite{Anderson2022}. Meanwhile, the SiC NV center retains the advantage of a hyperfine-coupled long-lived nuclear spin in the vicinity that can act as a quantum memory. However, none of these defects have demonstrated emission within the telecom range, and the search for further improved defects in this regard remains an ongoing endeavor.

To accelerate the search for advantageous point defects, high-throughput techniques have recently been applied to search the intrinsic defect space in SiC\cite{Davidsson2022} and extrinsic defect space in calcium oxide\cite{Davidsson2023}. It has been demonstrated that these can provide a detailed picture of possible defects and further explain observed properties, such as optical spectra. We have applied such high-throughput techniques for a large-scale study of extrinsic defects in SiC, consisting of $s$- and $p$-elements within single and double-defect clusters.
In this paper, we present the results from our high-throughput defect screening focusing on NV-like defects emitting in the near-telecom range, and provide a detailed characterization of the magneto-optical properties of the most interesting defect, the chlorine vacancy (ClV) center, which is noted in the screening as the most probable C-band emitter among the candidates.

The article is structured as follows. First, section \ref{sec:highthroughput} describes the high-throughput investigation performed in this work and summarizes the search result for diamond NV-like candidates in SiC. Section \ref{sec:methods} provides computational details of the ab-initio methods applied to the ClV defect, singled out as the most interesting candidate. In section \ref{sec:results_structure}, we introduce the defect, covering its properties related to geometric and electronic structure and its thermodynamical stability within 4H-SiC. In section \ref{sec:results_optical} and \ref{sec:results_phonons}, we summarize the most interesting optical properties of the defect, including both estimated brightness and electron-phonon coupling. In section \ref{sec:results_spin}, we summarize the ground state electron spin properties of the most stable spin-charge state. Finally, in sections \ref{sec:discussion} and \ref{sec:conclusions}, we discuss the prospects of the defect, given the predicted magneto-optical properties and similarity in electronic structure to other promising defects, and draw our conclusions.

\section{High-throughput investigations}\label{sec:highthroughput}

To efficiently search among feasible and interesting defects in SiC, we employ the high-throughput workflow package ADAQ\cite{Davidsson2021}, a collection of workflows designed for exhaustive defect generation and efficient point-defect characterization, which is implemented with the high-throughput toolkit \emph{httk}\cite{armientoDatabaseDrivenHighThroughputCalculations2020}. It is capable of high-throughput defect screening, in which possible charge and spin states are identified and analyzed for estimates of their optical properties. For a detailed description, see Ref.~\onlinecite{Davidsson2021}. The workflows perform density functional theory \cite{Kohn1965} (DFT) calculations using the Vienna Ab-initio Simulation Package\cite{Kresse1994,Kresse1996} (VASP), which uses plane-wave basis sets within the projector-augmented-wave method\cite{blochlProjectorAugmentedwaveMethod1994,kresseUltrasoftPseudopotentialsProjector1999}, with the Perdew-Burke-Erzenerhof\cite{Perdew1996} (PBE) functional.

The generated defect set consists of extrinsic $s$- and $p$-elements configured in single and double-defect configurations, considering combinations of substitutional and interstitial placements with possible vacancies. While it fully covers the extrinsic single-site defects and intrinsic-extrinsic pairs, it does not include every double-extrinsic pair due to the large amount of possible interstitial combinations. 
In total, 52\,600 defect configurations are considered in this search. However, defects sharing the same stoichiometry will likely have the most stable configuration realized at ambient conditions. Therefore, only defects lying on the defect hull\cite{Davidsson2021thesis,Davidsson2022}, i.e. being the most energetically favored within the same stoichiometry, were considered further.

Searching for defects with advantageous properties for quantum applications, we consider the diamond NV center as an ideal starting point. Following its favorable properties, we search the screened defects for spin triplet ground states, excluding single substitutionals and interstitials. Next, we narrow the search further by an upper bound on the formation energy ($<$ 6 eV), eliminating defects that are unlikely to form at high concentrations compared to other lattice defects, such as carbon and silicon vacancies. Finally, besides favorable spin properties and stability, the most interesting advantage would be emission in the telecom range. Hence, we finish the filtering by choosing defects with a zero-phonon line (ZPL) in the range of 0.5 eV to 1.1 eV, taking the screening accuracy into account\cite{Davidsson2018}. The final result is a group of five vacancy-substitution clusters, summarized in Table~\ref{tab:screening_result}.

\begin{table}[h!]
    \begin{minipage}{0.6\textwidth}
\begin{ruledtabular}
    \begin{tabular}{c|ccc|cc}
        Period & \multicolumn{3}{c|}{2} & \multicolumn{2}{c}{3}\\
        \hline
        Defect & V\textsubscript{Si}N\textsubscript{C} &
        V\textsubscript{Si}O\textsubscript{C} &
        V\textsubscript{Si}F\textsubscript{C} &
        V\textsubscript{Si}S\textsubscript{C} &
        V\textsubscript{Si}Cl\textsubscript{C} \\
        ZPL (eV) & 0.79 & 0.92 &  1.09 & 0.55 & 0.64  \\
        Dipole Moment (Debye) & 11.9 & 12.7 &    0.8 & 8.5 & 13.0\\
        Maximal Formation Energy (eV) & 4.52 & 2.22 & 3.23 & 4.87 & 5.85\\
    \end{tabular}
\end{ruledtabular}
    \end{minipage}
    \caption{The defects fulfilling all search criteria of a potential defect emitter with similarity to the diamond NV center. They were chosen based on having a maximal formation energy below 6 eV, a spin triplet ground state, and a ZPL from 0.5 eV to 1.1 eV. Elements are grouped based on the period of the extrinsic component.}
    \label{tab:screening_result}
\end{table}

We note the presence of the SiC NV center (V\textsubscript{Si}N\textsubscript{C}) among the filtered search results as a defect with a likeness to the NV center, known to be stable and with a strong luminescence. The inclusion of this defect in the search results confirms our choice of parameters, while it can act as a reference for the predicted quantities. Interestingly, the search results systematically consist of a silicon vacancy and dopant bonding on the carbon site. Bonding on the silicon site was observed among stable defects but with significantly higher formation energies. Furthermore, we observe a trend of increasing ZPL with each increasing element group within the same period. As the SiC NV center is known to emit in the infrared\cite{VonBardeleben2015}, defects in the same period are unlikely to yield transitions in the telecom region. However, turning to the third period clusters, the two defects with sulfur and chlorine components are shown to have ZPLs lower than the NV center and with the expected systematic screening error of approximately 0.2 eV\cite{Davidsson2018} would fall into the vicinity of 0.8 eV (1550 nm) where fiber-optic attenuation is minimal. Choosing between sulfur and chlorine clusters, the V\textsubscript{Si}Cl\textsubscript{C} features the largest dipole moment, implying higher radiative emission rates. Therefore, one can expect it to be comparably bright to the SiC NV center, if not brighter.

In summary, the screening yields several attractive candidates worthy of further investigation. In terms of bright telecom emission, we deem the V\textsubscript{Si}Cl\textsubscript{C} cluster, which we refer to as the chlorine vacancy in analogy to the NV center, to be the most promising candidate, and the remainder of this paper is primarily devoted to the detailed characterization of this defect.

\section{Characterization Methods}\label{sec:methods}
For the more detailed DFT calculations, we use a basis set with a 420 eV cutoff energy, applying the Heyd–Scuseria–\\Ernzerhof\cite{heydHybridFunctionalsBased2003} (HSE06) functional on 576 atom (6x6x2) 4H-SiC supercells at $\Gamma$-point, with a convergence threshold on the forces of 0.01 eV\AA\textsuperscript{-1}. Excited states are obtained according to the $\Delta$SCF approach\cite{Gorling1999,Kaduk2012}. Excited state structures are here converged to an energy difference of $5\cdot 10^{-5}$ eV.

The relative stability of the various charge and spin states is examined via the thermodynamic formation energies
\begin{equation}
    E^{q,S}_\text{form}(E_F) = E^{q,S}_\text{defect} - E_\text{bulk} - \sum_i n_i\mu_i + qE_\text{F} + E_\text{corr}^q\text{,}
\end{equation}
where $E^{q,S}_\text{defect}$ denotes the total energy of the defect state of (reduced) charge $q$ and spin state $S$, while $E_\text{bulk}$ is the energy of the defect-free host system. The $n_i$ and $\mu_i$ denote the change in chemical components to form the defect and respective chemical potential. The chemical potentials are here taken in the rich abundance limit, assuming the 0 K crystalline structures of each element. Finally, the electron potential energy of the charge state ($qE_F$) is added with a charge correction term $E_\text{corr}^q$ to account for finite size coloumbic effects. We employ the charge correction developed by Freysoldt et al.\cite{Freysoldt2009}.
Binding energies are calculated to provide a sense of the defect structural stability. Here, defect binding energies are defined as
\begin{equation}
    H^b[AB] = E_{\text{form}}[A] + E_{\text{form}}[B] - E_\text{form}[AB]\text{,}
\end{equation}
where the formation energy of the lowest lying defect state is considered for the defect decomposition $(\text{AB} \to \text{A} + \text{B})$. In evaluating the binding energies, we consider combinations of every stable component defect identified in the defect screening.

We treat the vibrational contribution to the photoluminescence according to Huang-Rhys theory of photon emission\cite{Huang2000,Tawfik2022}. Within this theory, we estimate the phononic coupling strength and the implied ratio of zero-phonon emission, in terms of the Huang-Rhys factor $S$ and Debye-Waller factor $W$. With the coupling strength to each phonon mode, we reconstruct the phonon sideband. In practice, we make use of the Pyphotonics toolbox\cite{Tawfik2022}.

The system phonon modes are obtained via the method of finite displacements.
However, as the defect significantly lowers the host crystal symmetry, complete phonon characterization may involve an excessive number of computations. Therefore, we perform the full set of phonon calculations on a single defect configuration of high symmetry. Otherwise, we employ a single-phonon approximation for an estimation of the average phonon coupling\cite{Alkauskas2012}. We use the Phonopy package\cite{Togo2015,Togo2023}, to construct displaced geometries, build the dynamical matrix of the system, and calculate the phonon eigenmodes. The PBE-functional is used within the phonon calculation framework, with the ground state structure converged to less than $10^{-7}$ eV\AA${}^{-1}$ in forces.

Properties of the defect electron spin is also examined if the concerned ground state hosts unpaired spins. In particular, we approximate the zero-field-splitting (ZFS) with the spin dipole-dipole interaction using the method in Ref.~\onlinecite{ZFSSensorPaper}, and we calculate the hyperfine structure of the lattice sites according to the method in Ref.~\onlinecite{Szasz2013}, as implemented in VASP.

 With a model Hamiltonian constructed with the spin parameters obtained from mentioned methods (see explicit formulation in Supplemental Materials\suppref), we investigate the coherence of the electron spin of the defect using the generalized cluster-correlation expansion (gCCE) approach, as implemented in the PyCCE toolbox\cite{Onizhuk2021}. We provide hyperfine values calculated as previously described, up to a distance of 12 \AA\ to the defect, including at least the third-nearest neighbor spins. For more distanced spins, we neglect the fermi-contact term and include only the dipole-dipole interaction term. The gCCE-method is applied using a maximum bath-spin distance of 50 \AA, and a maximum separation of cluster spins of 10 \AA. We use second order cluster expansion (gCCE-2) to obtain converged results and ensemble average over 250 bath spins configurations, each calculated with a Monte-Carlo bath state averaging using 100 bath states. 

\section{Results}
\subsection{Chlorine Vacancy Structure \& Ground States}\label{sec:results_structure}

\begin{figure}[h!]
    \centering
    \begin{minipage}{0.50\columnwidth}
        \tikz{
        \node [scale=1.5] at (-3.7,2.5) {\textbf{a)}};
        \node at (0,0) {\includegraphics[width=\textwidth]{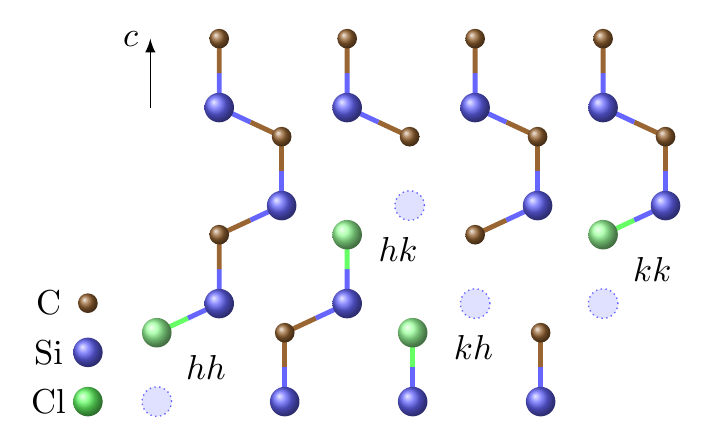}};
        }
    \end{minipage}
    \quad 
    \begin{minipage}{0.30\columnwidth}
        \tikz{
        \node at (0,0) {\includegraphics[width=\textwidth]{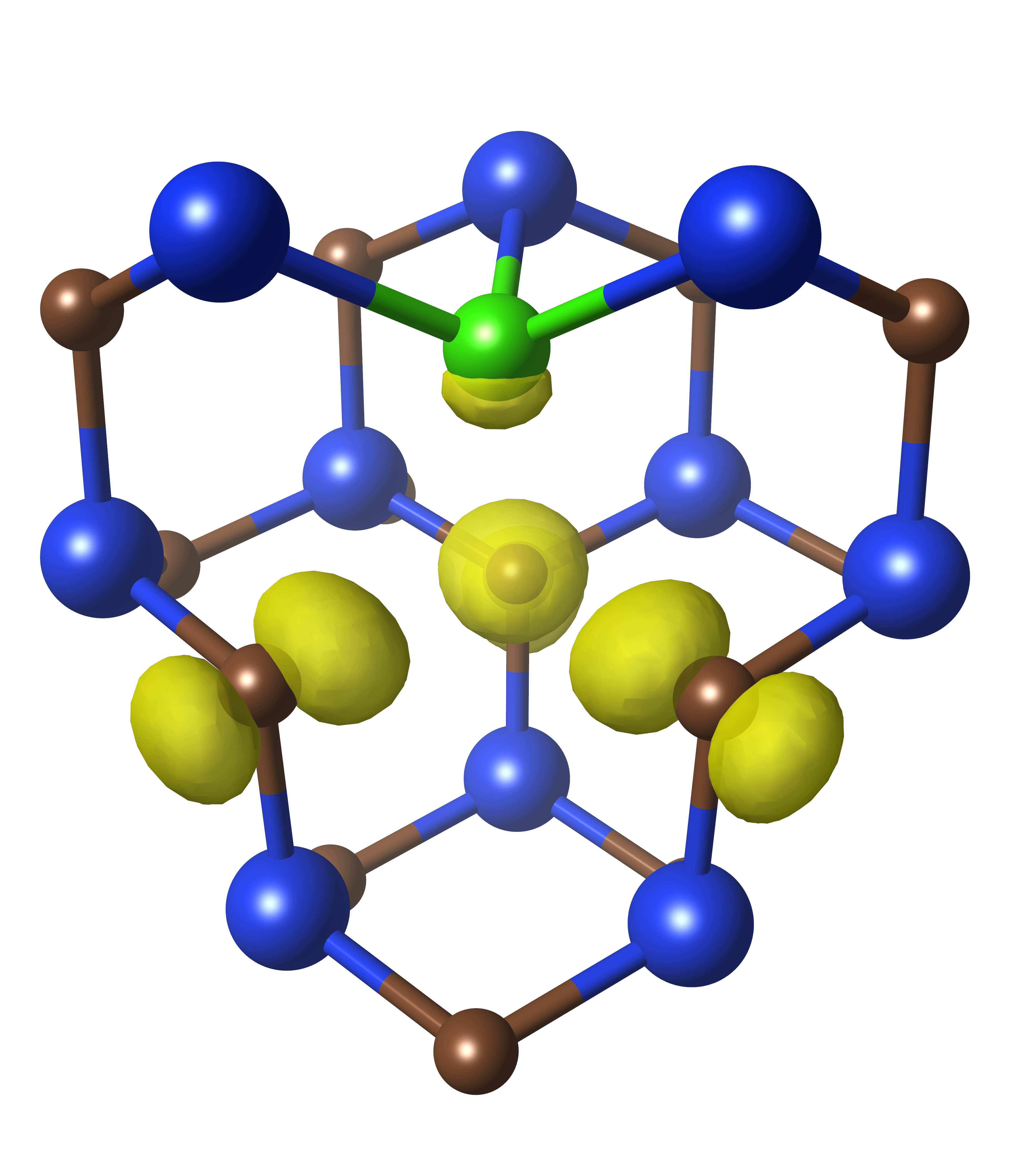}};
        \node [scale=1.5] at (-2.5, 2.5) {\textbf{b)}};
        }
    \end{minipage}
    \caption{The structure of the ClV defect in 4H-SiC. a) The four different nonequivalent defect configurations of 4H-SiC. b) The defect-local structure in the \emph{hh} configuration together with the $10^{-9}$ \AA${}^{-1}$ isosurface spin density of the occupied $a_1$ level.}
    \label{fig:structure}
\end{figure}

\begin{figure}[h!]
    \begin{center}
    \tikz{
    \node [scale=1.5] at (-5.7,2.5) {\textbf{a)}};
    \node at (-1,0) {
    \includegraphics[height=0.3\textwidth, clip, trim=1.50cm 0 0 0]{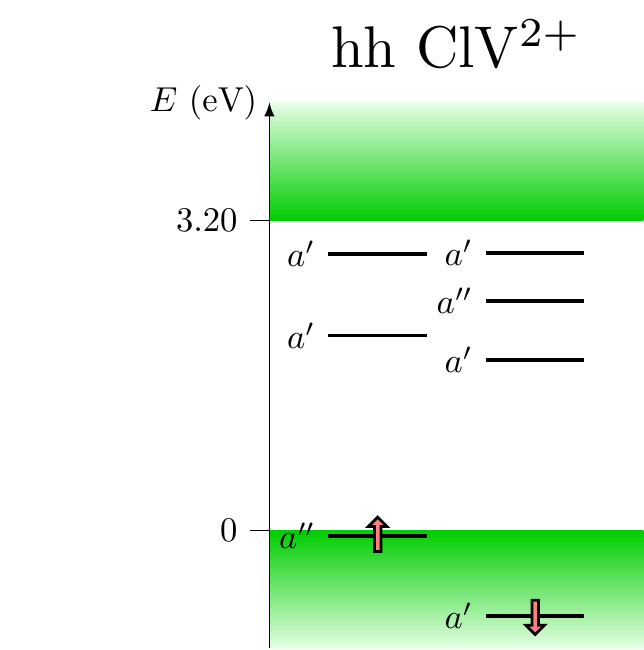}
    \includegraphics[height=0.3\textwidth, clip, trim=1.50cm 0 0 0]{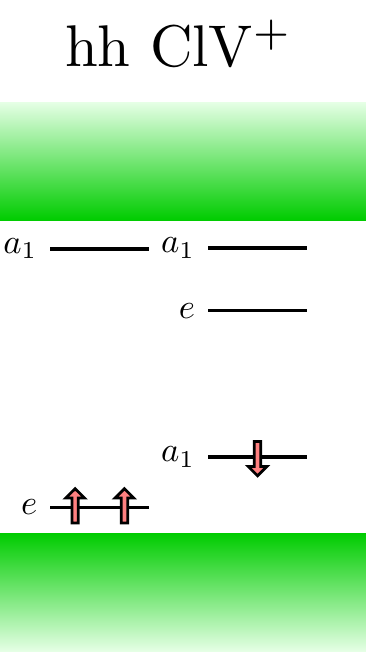}
    \includegraphics[height=0.3\textwidth, clip, trim=1.50cm 0 0 0]{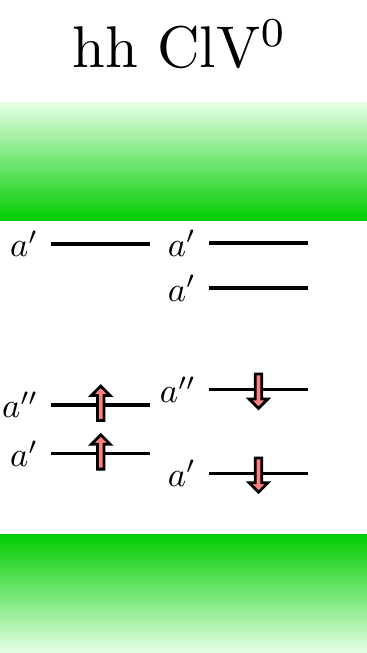}
    };
    \node [scale=1.5] at (6, 2.5) {\textbf{b)}};
    \node at (7,0) {\includegraphics[height=0.3\textwidth]{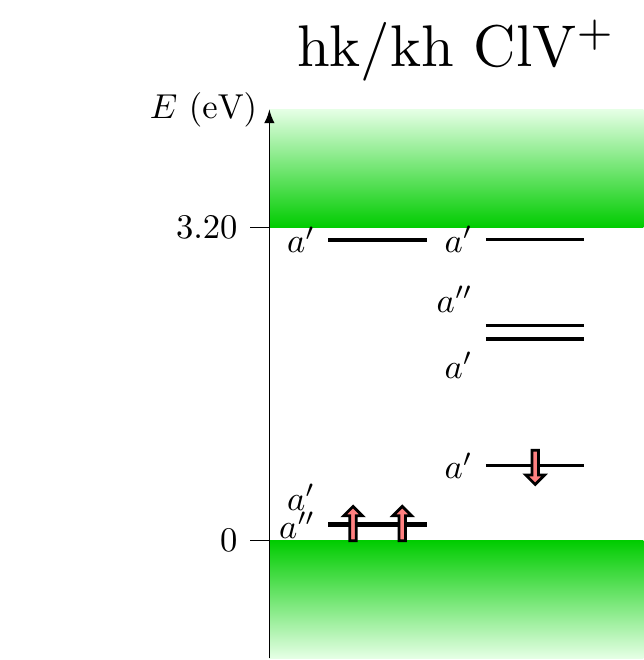}};
    }
    \end{center}
    \caption{The Kohn-Sham electronic band structure of relevant charge states, evaluated at the $\Gamma$-point. a) The band structure of double-positive, single-positive, and neutral charge states of the \emph{hh} configuration. b) The single-positive state band structure of the off-axis configurations, illustrating the band splitting due to the off-axis symmetry breaking.}
    \label{fig:band_structures}
\end{figure}

The ClV center consists of a silicon vacancy and a substitutional chlorine ion located on the neighboring carbon site. In 4H-SiC, a double-defect cluster is constructed in four nonequivalent lattice configurations labeled \emph{hh}, \emph{kk}, \emph{hk}, and \emph{kh} by the local environment of the defect sites (see Fig.~\ref{fig:structure}). The \emph{hh} and \emph{kk} configurations are on-axis, meaning the defect axis aligns with the \emph{c}-axis, making the defect C\textsubscript{3v} symmetric. The off-axis configurations lower the symmetry of the crystal further to C\textsubscript{1h}. The six dangling bonds of the silicon and carbon sites create localized defect states that appear in the band gap. For the most relevant states of the defect, the band gap states are for the on-axis (off-axis) configurations a lower $e$ state ($a'$ and $a''$ states) and a higher $a_1$ ($a'$) state in the spin-up channel, while in the spin-down channel an additional $a_1$ ($a'$) state is found to be the lowest-lying state in the gap. The only exception is the double-positive charge state, where the otherwise band-gap-present states are lowered into the valence band (see Fig.~\ref{fig:band_structures}). The other dangling bond states are found deep within the valence or conduction bands and are, therefore, not considered.

Fig.~\ref{fig:band_structures} shows the Kohn-Sham band structures and occupations of charge states, excluding the negative states. The \emph{hh} configuration is taken as representative of the defect. Note that the off-axis configurations will split the otherwise degenerate defect states due to their lower symmetry, as shown in Fig.~\ref{fig:band_structures}(b).
The most stable spin configurations are a triplet for the single-positive charge state and doublets for the double-positive and neutral charge states.
However, the single-positive state may also exhibit a meta-stable spin-singlet state, which by our calculations, lies 100 meV and 64 meV higher in energy for the on- and off-axis configurations, respectively.
The defect states are mainly centered on the neighboring carbons with a small contribution on the chlorine, as shown in Fig.~\ref{fig:structure}(b).
This qualitative localization feature is shared by both the SiC divacancy and NV center\cite{Ivady2016,Bockstedte2018}.

\begin{figure}[h!]
    \centering
    \includegraphics[width=0.4\columnwidth]{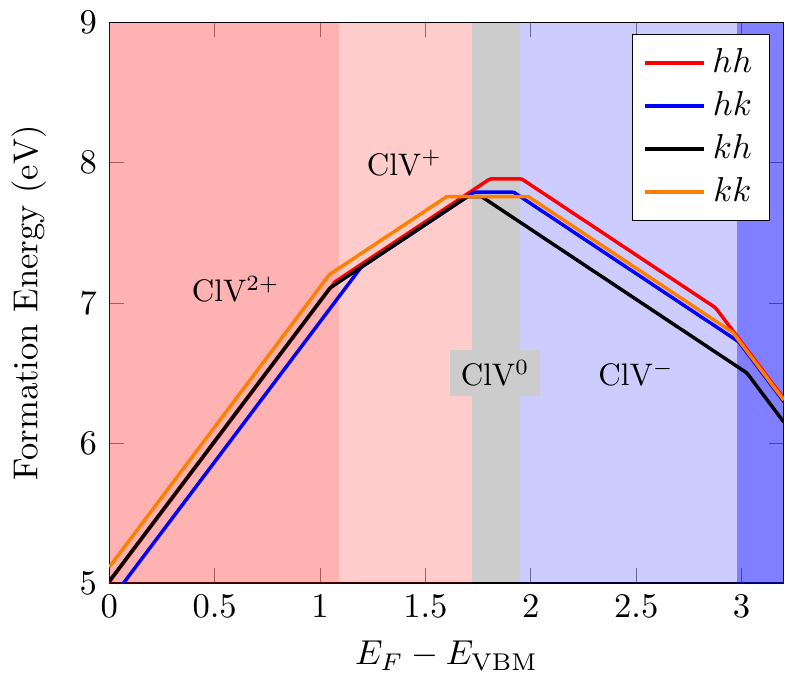}
    \caption{The formation energy of the lowest lying spin-charge states of the ClV defect. Colored regions mark the approximate region of relative stability of each state.}
    \label{fig:form_E}
\end{figure}

\begin{table}[h!]
    \centering
    \begin{minipage}{0.5\textwidth}
    \begin{ruledtabular}
    \begin{tabular}{c c c c c}
         State Transition & \emph{hh} & \emph{kk} & \emph{hk} & \emph{kh} \\
         \hline
        (2$+|+$) & 1.07 & 1.05 & 1.19 & 1.14 \\
        ($+|$0) & 1.76 & 1.73 & 1.73 & 1.69 \\
        (0$|-$) & 2.00 & 1.87 & 1.92 & 1.76 \\
        ($-|$2$-$) & 3.01 & 2.96 & 2.99 & 3.01
        \end{tabular}
    \end{ruledtabular}
    \end{minipage}
    \caption{The ionization levels of the ClV charge states, in eV.}
    \label{tab:ionization_levels}
\end{table}

Fig.~\ref{fig:form_E} shows the formation energy and Table~\ref{tab:ionization_levels} summarizes the ionization levels within the band gap.
Previous studies have considered additionally positive charge states of the defect on the level of the local density approximation\cite{Alfieri2012a}. However, these are likely to lack occupied defect orbitals in the band gap, hence are less interesting from the perspective of possible photon emitter states. In turn, the positive charge state, featuring several deep defect Khon-Sham levels in the band gap where optical transitions may occur, has a sizable stability region.
The (2$+|+$) varies from 1.05 to 1.19 eV $+$ $E_\text{V}$ and the ($+|$0) varies from 1.69 to 1.76 eV $+$ $E_\text{V}$.
In contrast, the neutral charge state is only stable in a smaller region between 1.69 to 1.92 eV $+$ $E_\text{V}$ and is greatly affected by the relative stability of the surrounding charge states, severely limiting this region of stability for the \emph{kh} configuration. 

\begin{figure}[h!]
    \centering
    \tikz{
    \node [scale=1.5] at (-3.3,3.1) {\textbf{a})};
    \node at (0,0) {\includegraphics[width=0.42\textwidth]{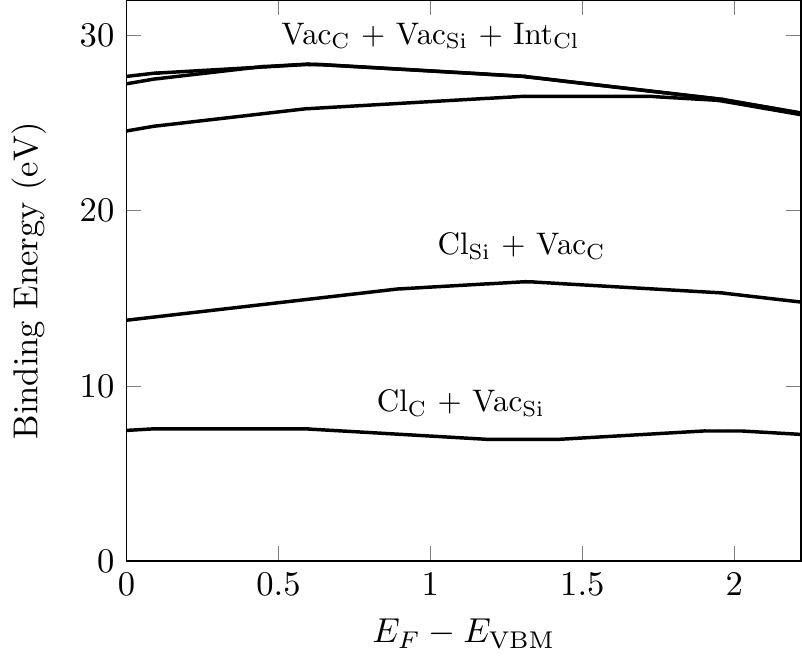}};
    }
    \tikz{
    \node [scale=1.5] at (-3.3,3.1) {\textbf{b})};
    \node at (0,0) {\includegraphics[width=0.42\textwidth]{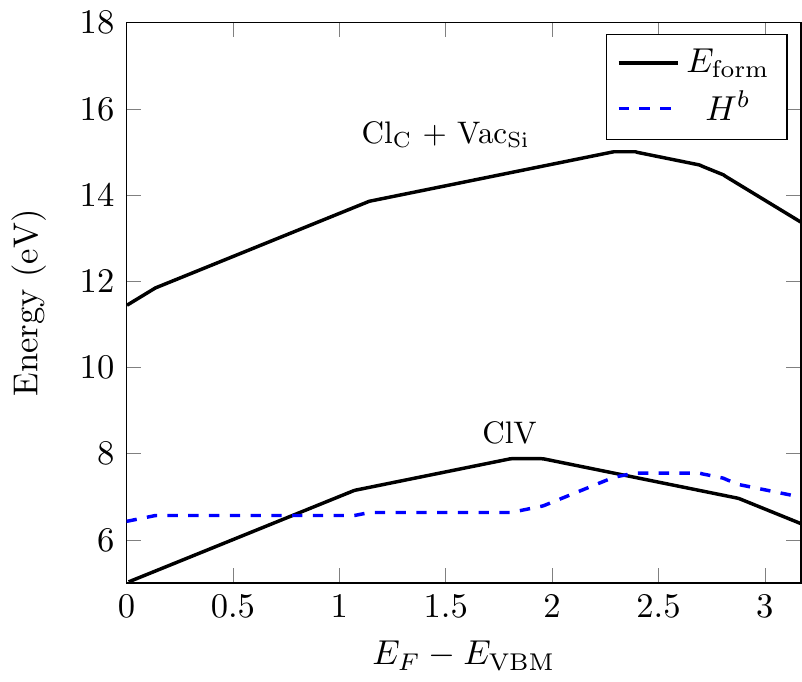}};
    }
    \caption{The binding energies of the ClV defect compared to possible single-defect separations, including three different identified chlorine interstitial sites, in the \emph{hh} configuration. a) The formation energy of all identified defect combinations and the ClV defect calculated with $\Gamma$-point PBE at screening accuracy. b) The formation energies, $E_\text{form}$, and binding energy, $H^b$, of the ClV and lowest-lying identified dissipation formation (separated chlorine substitution and silicon vacancy), calculated with the HSE06 functional.}
    \label{fig:binding_energy}
\end{figure}

The defect structure shown in Fig.~\ref{fig:structure}(a) is the lowest configuration for this stoichiometry. Notably, the energy difference (taken from screening data) between the studied structure and the relocation of the chlorine to the silicon site is about 4.45 eV.
Furthermore, the evaluation of binding energies to various component separations of the ClV center, using the screened defect data, shown in Fig.~\ref{fig:binding_energy}(a), indicates that the binding energy is greater than 7 eV.
For assurance of accuracy, the formation energy is evaluated for the lowest-lying separation with the HSE06 functional, which corresponds to the silicon vacancy separated from the chlorine substitution site.
As shown in Fig.~\ref{fig:binding_energy}(b), the change in binding energy from the PBE-predicted values is minimal.

\subsection{Optical Properties} \label{sec:results_optical}
\begin{figure}[h]
    \centering
    \includegraphics[height=0.35\textwidth, clip, trim=1.5cm 0 0 0]{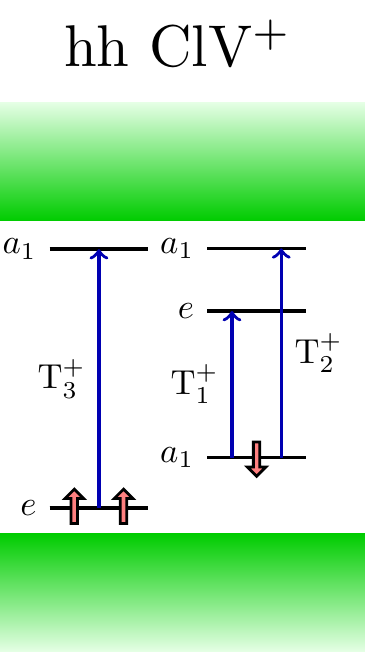}
    \includegraphics[height=0.35\textwidth, clip, trim=1.5cm 0 0 0]{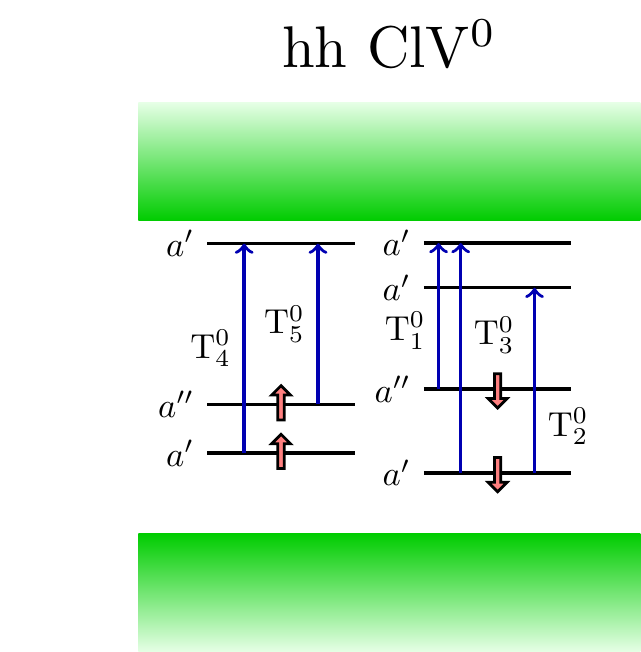}
    \caption{The \emph{hh} band structure with studied transitions marked, where $T^+_{\{1,2,3\}}$ and $T^0_{\{1,2,3,4,5\}}$ denote the studied transitions in the positive and neutral charge state respectively. The $T^+_1$, and other transitions involving a degenerate $e$ state, can consist of two transitions, either to the $a'$ or the $a''$ states of the $e$ state split by a Jahn-Teller distortion in the excited state or by structural symmetry reduction already present in the \emph{hk} and \emph{kh} configurations. However, we mainly consider the transition to the $a''$ state or the otherwise lowest-lying state.}
    \label{fig:transition_labels}
\end{figure}

Following the band structures of Fig.~\ref{fig:band_structures}, the defect-to-defect transitions exhibited in the positive and neutral charge states are labeled and illustrated in Fig.~\ref{fig:transition_labels}. We note that there is no label for the transition between the spin-down channel occupied $a''$ state to the lowest unoccupied $a'$ state in the neutral charge state.
These states originate from the same degenerate $e$ state from the positive charge state, which is split in the neutral charge state by the symmetry breaking Jahn-Teller effect.
Furthermore, we note that the positive charge state shares the electronic structure of the neutral divacancy and negative NV center\cite{Ivady2015a}.
The optical cycle of those defects involves the transition between the occupied $a_1$ state to the lowest lying unoccupied $e$ state\cite{Gali2010}. Here, we report on the corresponding transition in the ClV defect, which is denoted as $T^+_1$. As summarized in Table \ref{tab:transitions}, this transition is predicted to be particularly bright from the notably low radiative lifetime on the order of 30 ns, comparable to the predicted value for the diamond NV center of 6.5 ns using the same methodology\cite{Davidsson2023}. However, in contrast to the previously mentioned diamond and SiC defects which have been of high recent interest as single-photon emitters, this defect is predicted to emit in the telecom range with the ZPLs of the $T^+_1$ transition corresponding to wavelengths of 1330 nm, 1440 nm, 1490 nm and 1590 nm, respectively for the \emph{hh}, \emph{kk}, \emph{hk} and \emph{kh} configurations.
As is shown and predicted from the symmetry consideration of the transition, the emitted light will be polarized perpendicular to the $c$-axis in the on-axis configurations and otherwise likely to have mixed polarization. Additionally, we study two other transitions labeled $T^+_2$ and $T^0_2$ belonging to the positive and neutral ground state. Compared to the $T^+_1$ transition, these also feature low lifetimes, but in separate ZPL ranges. In particular, the $T^+_2$ is predicted to be in the visible spectrum while the $T^0_2$ is in the near-infrared region of 1100--1300 nm, most notable for being in the second biological window\cite{Smith2009}, where attenuation through blood and tissue is minimal. Transitions involving extended states in the valence or conduction band are not studied in detail here, but are estimated from the ionization levels indicated in Table~\ref{tab:ionization_levels} to have approximate bound-to-defect ZPLs of 1.7 eV and defect-to-band ZPLs of 2.1 eV.

\begin{table}[h!]
    \centering
    \begin{minipage}{0.49\textwidth}
    \begin{ruledtabular}
    \begin{tabular}{ccccc}
        Transition & Config. & \makecell{ZPL\\(eV)} & \makecell{Lifetime\\(ns)} & $\theta$\\ 
        \hline
        \multirow{4}{*}{T${}^+_1$} & $hh$ & 0.93 & 28.4 & $90.0^\circ$\\
        & $kk$ & 0.86 & 30.9 & $90.0^\circ$\\
        & $hk$ & 0.83 & 24.3 & $27.6^\circ$\\
        & $kh$ & 0.78 & 28.1 & $29.9^\circ$\\
        \hline
        
        \multirow{4}{*}{T${}^+_2$} & $hh$ & 1.64 & 63 & $0.1^\circ$\\
        & $kk$ & 1.74 & 40 & $0.5^\circ$\\
        & $hk$ & 1.74 & 35 & 67.2${}^\circ$\\
        & $kh$ & 1.76 & 81 & 79.1${}^\circ$\\

        \hline
        \multirow{4}{*}{T${}^0_2$} & $hh$ & 1.12 & 83.8 & $84.0^\circ$\\
        & $kk$ & 1.03 & 86.8 & $87.0^\circ$\\
        & $hk$ & 1.03 & 33.9 & $86.0^\circ$\\
        & $kh$ & 0.91 & 43 & $90.0^\circ$\\
    \end{tabular}
    \end{ruledtabular}
    \end{minipage}
    
    \caption{The properties of the particularly bright excited state transitions, for the positive and the neutral charge state. The data includes zero-phonon lines (ZPLs), the radiative lifetime of the state, and the polarization angle ($\theta$) between the transition dipole moment and the c-axis.}
    \label{tab:transitions}
\end{table}

\subsection{Electron-Phonon Coupling}\label{sec:results_phonons}
\begin{table}[h]
    \centering
    \begin{minipage}{0.5\textwidth}
    \begin{ruledtabular}
    \begin{tabular}{ccccc}
    Config. & $\Delta Q$ ($\text{amu}^{1/2}$\AA) & $\omega_g$ (meV) & $S$ & $W$\\
    \hline
    \emph{hh}   & 0.96  & 32.22 & 3.55 & 2.88\%\\
    \emph{kk}   & 1.02  & 30.82 & 3.85 & 2.12\%\\
    \emph{hk}   & 1.17  & 28.70 & 4.69 & 0.92\%\\
    \emph{kh}   & 1.11  & 29.84 & 4.43 & 1.19\%\\
    \end{tabular}
    \end{ruledtabular}
    \end{minipage}
    \caption{The phonon-coupling properties of the $T^+_1$ transition assuming a single-phonon approximation. Included is the single-phonon coordinate displacement, $\Delta Q$, the effective single-phonon frequency estimated from interpolated ground state energies, $\omega_g$, the obtained Huang-Rhys factor, $S$, and corresponding Debye-Waller factor,  $W$.}
    \label{tab:single-phonon results}
\end{table}

\begin{figure}[h!]
    \centering
    \includegraphics[width=0.5\textwidth]{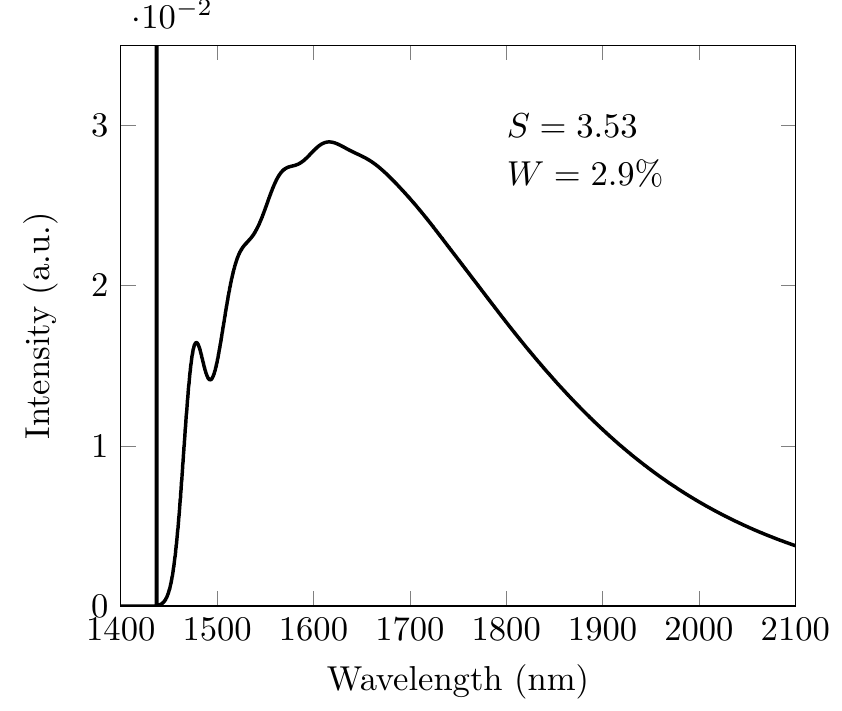}
    \caption{The phonon sideband of the $T^+_1$ transition for the \emph{hh} configuration, obtained after explicit phonon calculations and summation of partial phonon-contributions to the transition.
    The Huang-Rhys factor and corresponding quantum efficiency estimated from the multi-phonon coupling is also included.}
    \label{fig:sideband}
\end{figure}

We investigate the vibrational coupling of electron transition, focusing on the predicted telecom emission of the ClV$^+$ state. As explained in Section \ref{sec:methods}, calculated phonons of the defect ground state of the \emph{kk} configuration is used to qualitatively reconstruct the phonon sideband and quantify the overall coupling to phonons via Huang-Rhys theory. The calculated sideband in Fig.~\ref{fig:sideband} show discernible localized vibrational modes with a considerable contribution. Bumps in the spectrum can be associated with groups of local modes making up vibrational configurations with $a_1$ and $e$-state characteristics. In particular, the oscillation of the chlorine ion within the defect axis and perpendicular to the plane of reflection is a considerable component in the most strongly coupled modes.

As for the expected size of the optical phonon coupling, a Huang-Rhys factor of 3.53 was obtained for the \emph{kk} configuration corresponding to a Debye-Waller factor of 2.9\%. The zero-phonon contribution to the ClV luminescence is therefore noted to be lower than that of the diamond NV center (3.2\%\cite{Alkauskas2014}) and the neutral divacancy (5.3\%\cite{Christle2017}), which may be attributed to the relatively larger mass of the chlorine vacancy, as is majorly involved in the coupled vibrational modes. However, it should be noted that integrated photonics technologies available for SiC can increase this ratio considerably, with the divacancy demonstrated to achieve an enhanced Debye-Waller factor of more than 75\%\cite{Crook2020} with Purcell enhancement in a crystal cavity. Estimated quantities for all configurations are obtained in the single-phonon approximation and yield a quantum efficiency of approximately 1-3\% (see Table~\ref{tab:single-phonon results}), where the off-axis configurations are on the lower end of this estimate. However, it should be noted that compared with the multi-phonon estimation of the \emph{kk}, there is an overestimation of the Huang-Rhys factor in the single-phonon approximation. In fact, results show that the single-phonon approximation underestimates the contribution from low-frequency local modes at approximately 22 meV which are the dominating factor to the total Huang-Rhys factor (see Supplemental Material\suppref). On the other hand, the higher Huang-Rhys factors obtained in the off-axis configurations are most likely explained by an increased structural relaxation between ground and excited states, as indicated by the increased coordinate differences, $\Delta Q$.

\subsection{Spin Properties}\label{sec:results_spin}
\begin{table}[h!]
    \centering
    \begin{minipage}{0.5\textwidth}
    \begin{ruledtabular}
    \begin{tabular}{ccc}
        Config. & $D$ (GHz) & $E$ (MHz) \\
        \hline
        \emph{hh} & 1.93 & 0\\
        \emph{kk} & 1.87 & 0\\
        \emph{hk} & 1.90 & 209\\
        \emph{kh} & 1.82 & 150
    \end{tabular}
    \end{ruledtabular}
    \end{minipage}
    \caption{Zero-field splitting parameters of the ClV\textsuperscript{$+$} $S=1$ ground state. The $D$ and $E$ parameters are reported on the standard form of $D = \frac{3}{2}D_{zz}$ and $E = \frac{1}{2}(D_{xx} - D_{yy})$ with the principal components of the ZFS tensor $D_{xx}$, $D_{yy}$, $D_{zz}$.}
    \label{tab:ZFS_values}
\end{table}

We calculate the ZFS tensors of each ClV\textsuperscript{$+$} triplet configuration. Eigenvalues of the tensor are reported in Table~\ref{tab:ZFS_values}. We note the similarity in size of the $D$-components to those of the divacancy ($D\approx 1.3$ GHz)\cite{Koehl2011} and NV center ($D \approx 2.9$ GHz)\cite{Loubser1978}, which indicates that the ClV defect can be studied and used with spin-resonance techniques, such as optically detected magnetic resonance (ODMR), without considerable modification to experimental setups calibrated to the resonance of similar defects. One can also note the possibility of discerning the four configurations in spin-resonance-based techniques because of the observable variation in $D$.

\begin{figure}[h!]
\begin{minipage}{0.48\textwidth}
    \centering
    \tikz{
        \node [scale=1.5] at (-3.0,2) {\textbf{a})};
        \node at (0,0) {\includegraphics[width=0.8\textwidth]{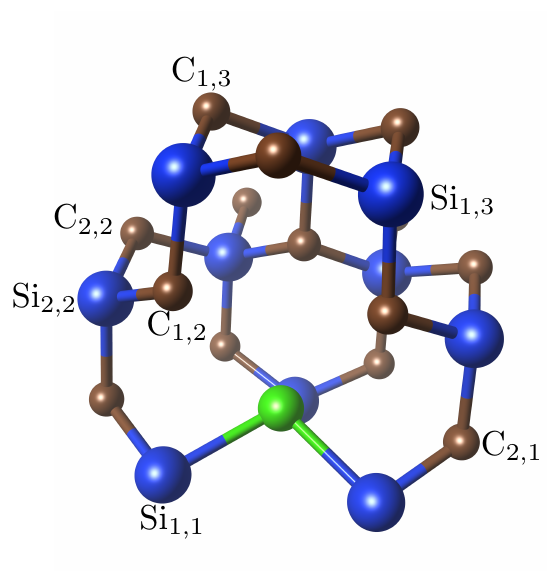}};
        \node at (0,-6) {\includegraphics[width=0.7\textwidth, clip, trim=1.3cm 0.5cm 1.4cm 0cm]{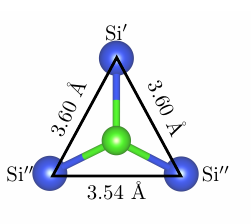}};
        \node [scale=1.5] at (-3.0,-5.5) {\textbf{b})};
    }
    \captionof{figure}{The \emph{hh} structure (a) with neighbor carbon and silicon sites, and the \emph{hk} structure (b) along the defect axis, showing the broken C\textsubscript{3v}. Labels follow the distance to the chlorine site, where the second index increases with distance along the defect axis. Symmetrically equivalent sites are not labeled, but primed quantities are introduced for the off-axis configurations to emphasize the reduced symmetry.}
    \label{fig:hyperfine_labels}
\end{minipage}
\begin{minipage}{0.48\textwidth}
    \centering
    \captionof{table}{The diagonalized hyperfine parameters (in MHz) for neighbor lattice sites of the ClV\textsuperscript{$+$} state. The diagonal components are sorted according to $|A_{xx}| > |A_{yy}| > |A_{zz}|$, and $\theta$ defines the angle between principal value $(A_{xx})$ and the defect symmetry axis. For second-nearest neighbor sites and beyond, we provide a range of the observed hyperfine values larger than 2 MHz.}
\begin{ruledtabular}
    \begin{tabular}{c  c r r r r }
       Ion & Config. & $A_{xx}$ & $A_{yy}$ & $A_{zz}$ & $\theta$\\
       \multirow{4}{*}{${}^{35}$Cl} & \emph{hh} & $-16.3$ & $-16.3$ & $-15.8$ & $90^\circ$\\
        & \emph{kk} & $-18.1$ & $-18.1$ & $-16.9$ & $90^\circ$\\
        & \emph{hk} & $-12.2$ & $-12.1$ & $-11.6$ & $85^\circ$ \\
        & \emph{kh} & $-9.9$ & $-9.3$ & $-9.0$ & $107^\circ$\\
        \hline
        \multirow{6}{*}{${}^{13}$C${}_{1,2}$} & \emph{hh} & 106 & 36.5 & 35.9 & 106${}^\circ$\\
         & \emph{kk} & 98 & 29.4 & 28.7 & 73$^\circ$\\
         & \emph{hk} (C$'$) & 102 & 39.4 & 38.4 & 105${}^\circ$\\
         & \emph{kh} (C$'$) & 96 & 36.8 & 36.3 & 106${}^\circ$\\
         & \emph{hk} (C$''$) & 107 & 34.4 & 33.8 & 107${}^\circ$\\
         & \emph{kh} (C$''$) & 98 & 30.1 & 29.6 & 106${}^\circ$\\
         \hline
        \multirow{6}{*}{${}^{29}$Si${}_{1,1}$} & \emph{hh} & 0.99 & 0.80 & 0.34 & 90${}^\circ$\\
         & \emph{kk} & 0.65 & 0.63 & $-0.06$ & 56${}^\circ$\\
         & \emph{hk} (Si$'$) & 0.58 & 0.56 & $-0.12$ & 90${}^\circ$\\
         & \emph{kh} (Si$'$) & $-1.08$ & 0.28 & 0.12 & 68${}^\circ$\\
         & \emph{hk} (Si$''$) & 0.60 & 0.49 & $-0.41$ & 105${}^\circ$\\
         & \emph{kh} (Si$''$) & 1.10 & 0.89 & 0.18 & 92${}^\circ$\\
         \end{tabular}
         \begin{tabular}{c c c c c}
         Ion & \multicolumn{4}{c}{$A_{xx},A_{yy},A_{zz}$}\\
          & \emph{hh} & \emph{kk} & \emph{hk} & \emph{kh}\\
         C\textsubscript{3,2} & 4.9 -- 10.5 & 5.4 -- 11.1 & 4.9 -- 10.6 & 5.0 -- 10.8\\
         C\textsubscript{1,3} & $|A| <$ 1 & $-8.5$ -- $-5.3$ & $|A| <$ 2 & $|A| < 1$\\
         C\textsubscript{2,3} & 3.2 -- 7.4 & 0.9 -- 2.7 & 3.3 -- 7.7 & 3.4 -- 8.6\\
         Si\textsubscript{2,2} & 9.4 -- 11.8 & 10.5 -- 12.4 & 8.7 -- 12.2 & 6.9 -- 13.8\\
         Si\textsubscript{4,2} & $|A| <$ 1 & $|A| <$ 1 & $-3.6$ -- $-0.5$ & $-3.3$ -- $-1.2$ \\
         Si\textsubscript{1,3} & 11.6 -- 12.1 & 12.8 -- 13.4 & 9.7 -- 12.7  & 9.9 -- 14.5\\
         Si\textsubscript{1,4} & $|A| <$ 2 & 2.4 -- 2.7 & $|A| <$ 2 & $|A| < 2$\\
         
    \end{tabular}
\end{ruledtabular}
    \label{tab:hyperfine_values}
\end{minipage}
\end{figure}

To determine the coupling of the defect electron spin to possible neighboring spins, the hyperfine structure is calculated at the neighboring lattice sites. These sites may feature ${}^{29}$Si and ${}^{13}$C doublet nuclear spins, which are found at respective natural abundances of 4.4\% and 1.1\%. Furthermore, the chlorine site of the defect will feature a quartet spin, but may be found as either ${}^{35}$Cl at 75.76\% natural abundance or ${}^{37}$Cl at 24.24\%. These sites are labeled as in Fig.~\ref{fig:hyperfine_labels}, where we number the sites according to rising position along, and distance to, the chlorine-to-vacancy axis. In the off-axis configurations, we distinguish between the lattice sites lying on and off the C\textsubscript{1h} plane of reflection, respectively denoted by primed and double-primed labels. In the on-axis configurations, a unique labeling along the defect axis is enough due to the rotational and reflection equivalence in the C\textsubscript{3v} symmetry. The hyperfine coupling may also differ between the two chlorine isotopes according to the difference in gyromagnetic ratio, which is reportedly 4.176 MHzT\textsuperscript{-1} for ${}^{35}$Cl and 3.476 MHzT\textsuperscript{-1} for ${}^{37}$Cl\cite{Harris2001}.

In Table~\ref{tab:hyperfine_values}, we report the results of the ClV$^+$ ground state, assuming a ${}^{35}$Cl isotope spin. Of particular note is the overall negativity of the chlorine principal components, indicating anti-parallel coupling of the defect triplet and nuclear quartet. Furthermore, the tensor is largely isotropic, but an order of magnitude larger than that of the nitrogen spin in the diamond NV center (approximately $-2$ MHz) \cite{Gali2008}. Otherwise, the couplings to nearest neighbor spins are quantitatively comparable to those reported from similar ab-initio considerations for the NV center in diamond and the divacancy in SiC\cite{Szasz2014}. We note the difference in coupling between carbon nearest neighbors and silicon neighbors. the coupling indicates a primary localization of the spin density toward the nearest neighbor carbons, in agreement with the illustrated orbital of Fig.~\ref{fig:structure}(b). Furthermore, as indicated by the hyperfine strengths of next-nearest neighbors, the spin density appears to extend perpendicular to the defect axis, coupling mainly to spins near the silicon vacancy site.

Using the obtained hyperfine and ZFS values, we calculate the coherence time, $T_2$, of the positive ClV center and obtain $T_2 \approx 1.45$ ms at an external field of 100 G. Once again, this is similar to what has been calculated and demonstrated for the divacancy\cite{Seo2016}, known to have one of the longest coherence times at natural abundance, and is close to the reported $T_2$ of the SiC host 1.1 ms\cite{Kanai2022}.

\section{Discussion}\label{sec:discussion}

We have shown that the ClV exhibit bright optical emissions and have predicted both qualitative and quantitative similarity to the diamond NV center. However, one consequence of these similarities that has not been examined in this paper is if there exists an optical cycle that populates one of the spin substates. The convenience of the solid-state qubits and emitters currently dominating the field is the presence of spin-dependent processes in excitation and state relaxation that allow for optical cycles that effectively polarize the spin state of the defect\cite{Bockstedte2018,Banks2019}. In both NV centers (in diamond and SiC) and the divacancy in SiC, these cycles have been enabled by a nonradiative intersystem crossing (ISC) from the excited triplet state onto an excited singlet state, which may de-excite onto the singlet ground state and transition into the triplet ground state through another ISC. The spin dependence of these processes finally favors the occupation of a particular spin state. The ClV defect also displays a triplet in the positive state and semi-stable singlet state that could accommodate a similar process. However, describing such a cycle requires insight into the effect of spin-orbit interaction on this defect, as well as the precise treatment of the singlet shelving states. The latter step is notoriously difficult as the excited shelving state displays high electron correlation and requires beyond-DFT methods to describe accurately\cite{Bockstedte2018}. Such detailed characterization is desirable but goes beyond the scope of the current work. However, assuming such a spin initialization cycle exists, the defect qubit techniques of similar defects could be generalized to the ClV defect. In addition, ODMR measurements could similarly enable the myriad of sensing schemes developed for the NV center. Given the ZFS parameters of the ClV electron spin, one may assume that ODMR would be applicable at a microwave drive of similar size as that of the NV center in diamond.

The most distinguishing feature of the ClV center is its bright emission in telecom. However, there are also transitions in the infrared and visible regions, which is rare in a single defect. However, the proximity of the visible excitation (ca. 1.7 eV) is considerably close to the predicted ($+|$0) ionization threshold and the accuracy in both ZPL and formation energies may be approximately 0.1 eV\cite{Deak2010,Komsa2012}. Therefore, it is difficult to say at this level of theory whether one can reach the higher $a_1$ state without considerable photoionization using a single-photon excitation. Interestingly, the difference in excitation energy between the $T^+_1$ and $T^+_2$ transitions is about 0.9 eV, which indicates that one could resonantly excite to the higher $a_1$ level via a two-photon process while avoiding photoionization.

Spin-to-charge conversion has been a particularly successful method of spin readout with majorly increased fidelity\cite{Zhang2021} based on distinguishing the spin state from spin-selective charge conversions of the center. In this context, we note the lack of likely localized defect transition in the ClV${}^{2+}$ state. Therefore, the double-positive state could be considered to be dark with respect to the predicted $T^+$ transitions, and a spin-to-charge readout to this state may be effective, assuming a spin-selective photoionization can be achieved and not suppressed by the ISC rate. Photoionization to the neutral state, as demonstrated for the divacancy\cite{Anderson2022}, can also be considered. However, based on the (2$+|+$) and ($+|$0) ionization levels in Fig.~\ref{fig:form_E} of ca. 1 eV and 1.7 eV from the valence band, photoionization with a two-photon process will likely require secondary photon energies higher than for the initial excitation wavelength at telecom energies.

Our spin property characterization clearly illustrates the similarity to the divacancy and NV centers. Regarding hyperfine coupling to lattice spins, the dominating source of spin decoherence at natural abundance in SiC, there is close to no quantitative difference, and we expect coherence times to coincide apart from the effect of the present chlorine spin. However, as indicated by the calculated $T_2$ time of 1.45 ms, this defect can exhibit long coherence times at natural abundance, shown to be close to or longer than the coherence time of the divacancy, having the longest coherence time at natural abundance among the known defects in SiC\cite{Seo2016}. The longer coherence times could be attributed to the chlorine spin coupling, adding an antiparallel contribution to the otherwise spin-parallel mean field. 
These results are mainly valid at $T=0$ K, but the temperature effects on the defect are vital to its applicability as a qubit. The room-temperature properties of the diamond NV center have been attributed to low spin-orbit interaction and high Debye temperature of diamond\cite{Seo2016}, leading to millisecond room-temperature coherence times\cite{Herbschleb2019}. While we do not include temperature effects in this work, we note that room-temperature coherent control has been demonstrated for both the divacancy\cite{Li2022} and the SiC NV center\cite{Wang2020}, achieving a room-temperature coherence time of 24 $\mu$s. Given the ClV likeness to these defects and calculated $T_2$ time in this work, it is plausible that the same room-temperature properties can be achieved.

Finally, having shown the potential properties of this highly interesting defect, the authors note that the ClV and other chlorine-related centers have been largely unexplored in the field of quantum technology. Despite chlorine being a central component in CVD growth of SiC crystals, there is a lack of publications reporting chlorine centers with notable photoluminescence. This may indicate that, despite chlorine and intrinsic defects being present during the crystal growth process, a barrier or competing process prevents the formation of the stable defect cluster.  However, deep-level transient spectroscopy measurements at high annealing temperatures in chlorine-implanted p-type SiC have observed defect levels attributed to chlorine-related centers. Notably, Ref.~\onlinecite{Alfieri2012} predicts one such charge level of $E_\textsubscript{V} + 0.97$ eV that agrees with the ($2+|+$) level of the ClV center presented in this paper. This results suggests that chlorine needs to be implanted to create the ClV center.

\section{Conclusions}\label{sec:conclusions}
In this work, we have done a large-scale search of the extrinsic defect space within 4H-SiC using the workflow package ADAQ, identifying four defects that are worthy of further investigation: The V\textsubscript{Si}O\textsubscript{C}, V\textsubscript{Si}F\textsubscript{C}, V\textsubscript{Si}S\textsubscript{C} and ClV centers. We report in detail on the most intriguing defect, the ClV center, whose predicted optical properties are promising for a future single-photon emitter.
Our hybrid-DFT calculations predict strong thermodynamic stability and bright emission within three optical regions of major interest in single-photon nano-devices, the visible, infrared, and telecom regions. Meanwhile, it shares many structural and electronic state properties with defects that have dominated the literature over the last decade. While its natural quantum efficiency is on par with the NV center, it is offered greater opportunity for enhancement using the SiC mature fabrication techniques and up-and-coming nanophotonics technology. The overarching similarity between the ClV and other known defects implies it should hold the properties of a state-of-the-art solid-state qubit, with optical spin initialization and readout.
Simultaneously, it exhibits noteworthy coherence qualities. Apart from the presence of a quartet nuclear spin in its vicinity for state storage and control, the triplet electron spin coherence time is found to be one of the longest in natural abundance SiC. Further investigation into its excited state spin properties and external field sensitivity will tell more of its prospects for quantum sensing. The ClV center has the qualities which made the diamond NV center a focal point of solid-state quantum technologies. As it further addresses the demand for telecom range emission, it offers an improved defect candidate for future quantum technologies.

\begin{acknowledgments}
We acknowledge support from the Knut and Alice Wallenberg Foundation (Grant No. 2018.0071). Support from the Swedish Government Strategic Research Area Swedish e-science Research Centre (SeRC) and the Swedish Government Strategic Research Area in Materials Science on Functional Materials at Linköping University (Faculty Grant SFO-Mat-LiU No. 2009 00971) are gratefully acknowledged. JD and RA acknowledge support from the Swedish Research Council (VR) Grant No. 2022-00276 and 2020-05402, respectively. We acknowledge PRACE for awarding us access to MareNostrum at Barcelona Supercomputing Center (BSC), Spain. The computations were also enabled by resources provided by the National Academic Infrastructure for Supercomputing in Sweden (NAISS) and the Swedish National Infrastructure for Computing (SNIC) at NSC and PDC partially funded by the Swedish Research Council through grant agreements no. 2022-06725 and no. 2018-05973.
\end{acknowledgments}



%

\end{document}


\title{Supplemental Materials}
\maketitle

\section{Defect Characterization Methods}
\subsection{State Locality}
With the converged ground states, the Kohn-Sham orbital states are analysed to identify defects states and possible vertical electronic transitions. Defect states are identified by their locality, using the inverse participation ratio (IPR) of the wavefunction $\phi$ as a locality measure
\begin{equation}
    \operatorname{IPR}(\phi) = \frac{\int_V |\phi(\vec{r})|^4 \dd V}{\int_V |\phi(\vec{r})|^2\dd V}\text{,}
\end{equation}
where integration is performed over the supercell volume $V$. Following this measure, the delocalized band states will yield an IPR value close to $1/V$, while defect states will upon increasing locality approach $\text{IPR} = 1$. The IPR measure may also be applied to phonon states in an analogous manner for the dislocations $\vec{u}_{\lambda,\alpha}$, with mode index $\lambda$ and atom index $\alpha$,
\begin{equation}
    \operatorname{IPR}_\lambda = \frac{\sum_\alpha |\vec{u}_{\lambda,\alpha}|^2}{\sum_\alpha |\vec{u}|}.
\end{equation}

\subsection{Transition Properties}
Properties of the excitated states are obtained by comparing the excited state to the corresponding ground state. In particular, the ZPL of the optical transition is calculated as the difference in total energies of the relaxed excited and ground states,
\begin{equation}
    E_{\text{ZPL}} = E_\text{ex,min} - E_\text{g,min}\text{.}
\end{equation}
We also analyze the strength of the transition, characterized by the transition dipole moment\cite{Davidsson2020}, here defined as
\begin{equation}
    \vec{\mu}_{i,f} = \bra{\phi_f}e\vec{r}\ket{\phi_i}\text{,}
\end{equation}
where $\ket{\phi_i}$ and $\ket{\phi_f}$ are taken as the ground state initial state orbital and the excited final state orbital, respectively. One can accordingly estimate the low-temperature radiative lifetime of the transition via Fermi's golden rule\cite{Wu2019}, arriving at the expression
\begin{equation}\label{eq:lifetime}
    \tau_{i,j} = \frac{3\varepsilon_0 \hbar c^3}{16n\pi^3\nu^3|\vec{\mu}_{i,j}|^2}\text{,}
\end{equation}
where $n$ is the refractive index of the host at the relevant wavelength, here taken as 2.6473, and $\nu$ being the corresponding frequency of the ZPL.

\subsection{Spin Model}
A model Hamiltonian is made using the obtained zero-field-splitting and hyperfine tensors, on the form,
\begin{equation}
    \mathcal{H} = \gamma_e B S_z + \sum_{i=1}^N\gamma_i \vec{B} \cdot \vec{I}_i + D\left(S_z^2 - \frac{1}{3}S(S+1)\right) + E(S_x^2 - S_y^2)+ \sum_{i=1}^N \vec{S}\tensor{A_i}\vec{I}_i + \sum_{i<j}^N \vec{I_i} \tensor{P}_{ij} \vec{I}_j,
\end{equation}
describing a linear Zeeman effect of electron spin, $\vec{S}$, and the $N$ nuclear spins, $I_i$, with respective gyromagnetic ratios $\gamma_e$ and $\gamma_i$ and a magnetic field $B$ fully aligned with the quantization axis of the electron spin; zero-field splitting, hyperfine coupling $\tensor{A_i}$ between defect and nuclear spins, and nuclear spin interactions $\tensor{P}_{ij}$ here taken to be fully dipolar. Gyromagnetic ratios for the defect electron spin is assumed not to deviate from that of a single electron spin. Nuclear spin gyromagnetic ratios are taken as tabulated in Ref.~\onlinecite{Harris2001}. We note that dipolar interaction terms with spins from dopants and other defects are not considered, and only the interaction with nuclear spins is investigated here.

Having calculated the time evolution of the coherence function $\mathcal{L}(t)$ of the defect electron spin using the generalized cluster-correlation expansion, we fit a squeezed exponential to the decay envelope of the coherence function, for extraction of the $T_2$ coherence time,
\begin{equation}
    \mathcal{L}(t) = e^{-\left(t/T_2\right)^n}.
\end{equation}

\subsection{Phononic Contribution to Luminescence}
Following Huang-Rhys theory of photoluminescence, the initial transition state is taken to linearly couple to a set of vibrational states approximated as harmonic oscillator states. The strength of the coupling between transition and the vibrational states is quantified by the Huang-Rhys factor
\begin{equation}
    S = \sum_\lambda S_\lambda = \sum_\lambda \frac{\omega_\lambda (\Delta Q_\lambda)^2}{2\hbar}\text{,}
\end{equation}
with partial Huang-Rhys factors $S_\lambda$ for the coupling to each vibrational mode, determined by the mode frequency $\omega_\lambda$ and the overall change in configuration coordinates along the mode
\begin{equation}
    \Delta Q_\lambda = \sum_a \sqrt{m_a}(\vec{R}_{\text{ex},a} - \vec{R}_{\text{g},a})\cdot \hat{u}_{\lambda,a},
\end{equation}
where the sum is over lattice sites with mass $m_a$, lattice positions in the excited ($\vec{R}_{\text{ex},a}$) and ground state ($\vec{R}_{\text{g},a}$), and the unit vibrational displacement of the mode $\hat{u}_{\lambda,a}$. The ratio between zero-phonon and phonon-coupled emission, is then given by the Debye-Waller factor
\begin{equation}
    W = e^{-S}\text{.}
\end{equation}
In the case of a single-phonon approximation of the excited state vibrational coupling, the Huang-Rhys factor reduces to
\begin{equation}
    S = \frac{\omega_g(\Delta Q)^2}{2\hbar},
\end{equation}
with $\omega_g$ being the effective phonon frequency extracted from fitting a parabola to the energy of interpolated structures between ground and excited state, and the coordination coordinate distance expressed as
\begin{equation}
    \Delta Q^2 = \sum_a m_a \left|\vec{R}_{\text{ex},a} - \vec{R}_{\text{g},a}\right|^2\text{.}
\end{equation}
Within the same theory, the phonon sideband for a transition between state $i$ and $j$ may be obtained as,
\begin{equation}
    I_{i,j}(\omega) = I_0 |\vec{\mu}_{i,f}| \sum_{\lambda=1}^M e^{-S}\frac{S^\lambda}{\lambda!}\delta(E_{\text{ZPL}} - \hbar \omega_\lambda - \hbar\omega),
\end{equation}
with some intensity scaling factor $I_0$, summing over all $M$ vibrational modes.

\newpage

\section{Optical Characterization Using ADAQ}

\begin{figure}[h!]
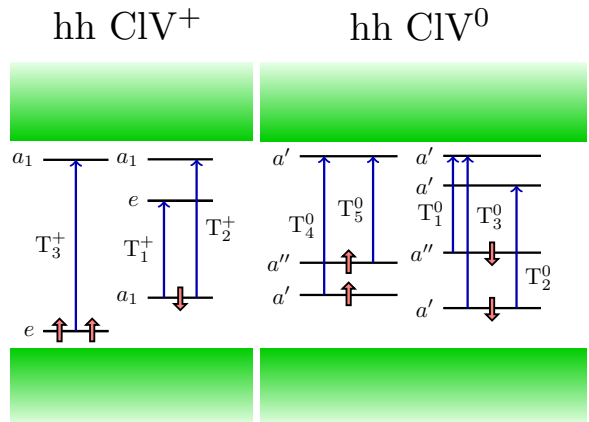

    \centering
    \includegraphics[height=0.32\textwidth, clip, trim=1.5cm 0 0 0]{main_figures/hh_ClVp_transitions}
    \includegraphics[height=0.32\textwidth, clip, trim=1.5cm 0 0 0]{main_figures/hh_ClV0_transitions}
    \caption{The \emph{hh} band structure with studied transitions marked, where $T^+_{\{1,2,3\}}$ and $T^0_{\{1,2,3,4,5\}}$ denote the studied transitions in the positive and neutral charge state respectively. }
    \label{fig:transition_labels}
\end{figure}

The defect excitations were treated at several levels of theory. In the screening workflow, less converged results are obtained with the PBE functional at $\Gamma$-point. Interesting defects are further investigated at higher convergence with a 2x2x2 k-point sampling of the Brillouin zone within an ADAQ workflow fully exploring all possible excitations. Finally, hybrid-DFT methods are applied as explained in the main text. However, not all transitions illustrated in Fig.~\ref{fig:transition_labels} are investigated at the level of hybrid-DFT because of the computational cost. Therefore, in Table \ref{tab:ADAQ_transitions} we present all converged excitation data calculated from the detailed ADAQ workflow. Not all transitions are included however, as ADAQ attempts to reduce the effort wasted on states that do not converge and disregards excited states that show signs of difficult convergence.

\begin{table}[h!]
    \centering
    \begin{minipage}{0.45\textwidth}
    {\Large a)}
    \vspace{2mm}
    \begin{ruledtabular}
    \begin{tabular}{ccccr}
        Transition & Config. & \makecell{ZPL\\(eV)} & \makecell{Dipole Moment $|\vec{\mu}|$\\(Debye)} & $\theta$\\ 
        \hline
        \multirow{4}{*}{T${}^+_1$} & $hh$ & $0.76$ & $8.9$ & $86.6^\circ$\\
        & $kk$ & $0.72$ & $9.1$ & $88.1^\circ$\\
        & $hk$ & 0.70 & 12.9 & $27.3^\circ$\\
        & $kh$ & 0.65 & 13.4 & $24.1^\circ$\\
        \hline
        
        \multirow{4}{*}{T${}^+_2$} & $hh$ & 1.62 & 3.23 & $0.60^\circ$\\
        & $kk$ & 1.74 & 3.29 & $3.71^\circ$\\
        & $hk$ & 1.73 & 2.74 & 70.1${}^\circ$\\
        & $kh$ & 1.78 & 1.55 & 73.7${}^\circ$\\

        \hline
        \multirow{4}{*}{T${}^+_3$} & $hh$ & 1.44 & 1.21 & $86.9^\circ$\\
        & $kk$ & 1.56 & 0.43 & $79.1^\circ$\\
        & $hk$ & 1.57 & 0.94 & $78.1^\circ$\\
        & $kh$ & 1.63 & 0.59 & $65.7^\circ$\\
    \end{tabular}
    \end{ruledtabular}
    \end{minipage}
    \quad
    \begin{minipage}{0.45\textwidth}
    {\Large b)}
    \vspace{2mm}
    \begin{ruledtabular}
    \begin{tabular}{ccccr}
        Transition & Config. & \makecell{ZPL\\(eV)} & \makecell{Dipole Moment $|\vec{\mu}|$\\(Debye)} & $\theta$\\ 
        \hline
        \multirow{4}{*}{T${}^0_1$} & $hh$ & 0.81 & 2.29 & $87.8^\circ$\\
        & $kk$ & 0.97 & 0.86 & $81.3^\circ$\\
        & $hk$ & 0.99 & 2.17 & $78.5^\circ$\\
        & $kh$ & 1.08 & 0.90 & $52.6^\circ$\\
        \hline
        
        \multirow{4}{*}{T${}^0_2$} & $hh$ & 0.78 & 9.24 & $88.7^\circ$\\
        & $kk$ & 0.72 & 9.37 & $89.1^\circ$\\
        & $hk$ & 0.71 & 11.5 & 75.1${}^\circ$\\
        & $kh$ & 0.67 & 12.2 & 85.6${}^\circ$\\

        \hline
        \multirow{4}{*}{T${}^0_3$} & $hh$ & 1.59 & 3.45 & $5.69^\circ$\\
        & $kk$ & 1.70 & 3.59 & $3.94^\circ$\\
        & $hk$ & 1.68 & 3.01 & 74.7${}^\circ$\\
        & $kh$ & 1.74 & 1.79 & 74.1${}^\circ$\\

        \hline
        \multirow{4}{*}{T${}^0_4$} & $hh$ & $\ast$ & $\ast$ & $\ast$\\
        & $kk$ & 1.08 & 0.64 & $56.2^\circ$\\
        & $hk$ & 1.09 & 0.96 & $68.6^\circ$\\
        & $kh$ & 1.19 & 0.73 & $23.1^\circ$\\
    \end{tabular}
    \end{ruledtabular}
    \end{minipage}
 
    \caption{The properties of the excited state transitions for the positive and the neutral charge state, as calculated by ADAQ with the PBE functional with a 2x2x2 k-point grid. The data includes zero-phonon lines (ZPLs), the transition dipole moment, and the polarization angle ($\theta$) between the transition dipole moment and the c-axis. Table a) shows the positive state transitions $T^+_{1,2,3}$, and b) shows the neutral state transitions $T^0_{1,2,3,4,5}$. Starred ($\ast$) entries were not fully converged by the ADAQ workflow, due to exhibiting signs of difficult convergence. For the same reason, transition $T^0_5$ is not included in the data set. }
    \label{tab:ADAQ_transitions}
\end{table}


%